\documentclass[showpacs,preprintnumbers,amsmath,amssymb,aps]{revtex4}

\usepackage{graphicx}
\usepackage{dcolumn}
\usepackage{bm}

\begin{document}

\preprint{}

\title{Kinetics of propagating phase transformation in compressed bismuth}

\author{Marina Bastea}
\email{bastea1@llnl.gov}
\author{Sorin Bastea}
\author{James A. Emig}
\author{Paul T. Springer}
\author{David B. Reisman}
\affiliation{Lawrence Livermore National Laboratory, P.O. BOX 808, Livermore, CA 94550}


\begin{abstract}
We observed dynamically driven phase transitions in isentropically compressed 
bismuth. By changing the stress loading conditions we explored two distinct cases: 
one in which the experimental signature of the phase transformation corresponds to 
phase-boundary crossings initiated at both sample interfaces, and another 
in which the experimental trace is due to a single advancing transformation front 
in the bulk of the material. We introduce a coupled 
kinetics - hydrodynamics model that for this second case enables us, 
under suitable simplifying assumptions, to directly extract characteristic 
transition times from the experimental measurements.
\end{abstract}


\maketitle

The kinetics of first-order phase transformations has long been a topic of great 
experimental and theoretical interest \cite{gms83,lt73}. Phase separation is, for example, 
a common technologically important occurrence in many alloys \cite{fpl99}, 
while structural-transition kinetics is believed to be relevant for understanding the 
dynamics of Earth's mantle \cite{ss94}. The development of high pressure experimental 
techniques has brought new perspectives on this problem, and new insights on long-standing 
scientific puzzles, e.g. the formation of natural diamond \cite{ouk04}. 
Understanding the kinetics of high pressure phase 
transitions is also an important step in fulfilling the promise of high pressure science 
to help develop new materials for technological applications \cite{pm02}. Dynamic compression 
experiments allow the study of such non-equilibrium processes occurring on very short timescales 
- $10^{-12}$ to $10^{-6} s$, which are otherwise difficult to investigate with traditional, static 
high pressure techniques. We present here the results of isentropic compression experiments 
exploring non-equilibrium behavior associated with polymorphic phase transitions in bismuth.

\begin{figure}
\includegraphics{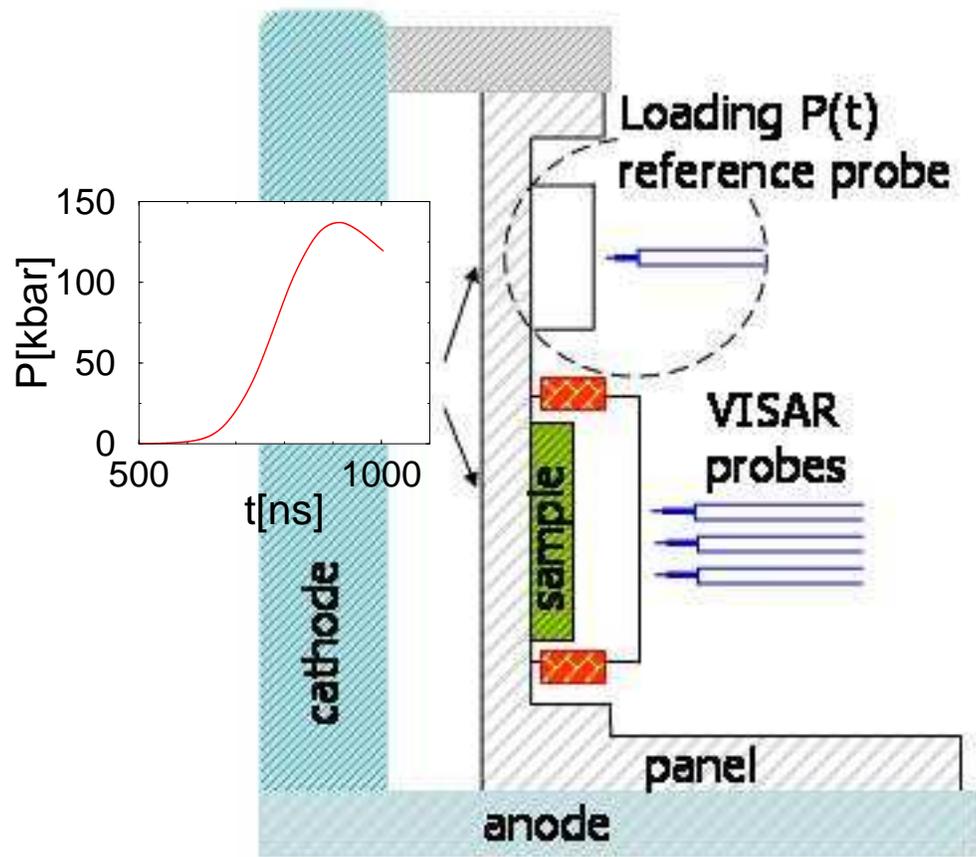}
\caption{Schematic cross-section through target assembly. Bi sample is contained between 
the panel (Cu or Al) and the transparent window (LiF or Sapphire). Heat is applied to the 
sample through a band heater (red) wrapped around the circumference of the window. A 3 mm 
vacuum gap (AK gap) exists between the panel and cathode. A rapidly varying magnetic field 
in the AK gap generates the pressure pulse that compresses the Bi sample. A reference probe 
assembly consisting of a transparent window impedance matched and glued to the panel provides 
a direct measurement of the loading pressure profile for each sample - see inset P(t).}
\label{target}
\end{figure}

The experiments were carried out using high purity poly-crystalline bismuth samples shaped 
as disks with $8-10 mm$ diameter and $0.3-0.6 mm$ thickness, with very flat and parallel 
surface finish obtained by diamond turning. The initial conditions were ambient pressure and 
temperatures between $\simeq 300K$ and $\simeq 400K$, where bismuth has a well studied rhombohedral 
crystal structure - Bi(I). We applied a smooth, magnetically driven pressure ramp to the 
target containing the sample - see Fig. \ref{target} for the experimental set-up, with 
duration of $\simeq 300 ns$ and $\simeq 150 kbar$ maximum value. As a result the system 
was driven along a quasi-isentropic thermodynamic path that first crosses the Bi(I) phase boundary 
into the Bi(II) phase, with a centered monoclinic crystal structure. We measured the time 
dependence of the velocity of the interface between the sample and a transparent window using 
velocity interferometry technique (VISAR) \cite{VISAR}. The loading pressure was carefully 
designed to avoid developing shocks in the sample before the phase transformation conditions 
were achieved, and monitored in each experiment with a reference probe. The details of the 
magnetic pulse generation are similar with the ones described in \cite{ice-tech}. To insure 
high accuracy results the initial temperature variation across the sample diameter was 
continuously monitored, and was found to be $\leq 5K$. Also, the velocity of the sample/window 
interface was measured on several points, spaced up to 2 mm apart, each traced with 1-2 
interferometers with different sensitivities to eliminate fringe loss uncertainties. The 
windows used in the experiments were [100] single crystal lithium fluoride - $LiF$ and 
sapphire. Their optical properties in the pressure-temperature regime accessed in these 
experiments are summarized in \cite{win1,win2}.

\begin{figure}
\includegraphics{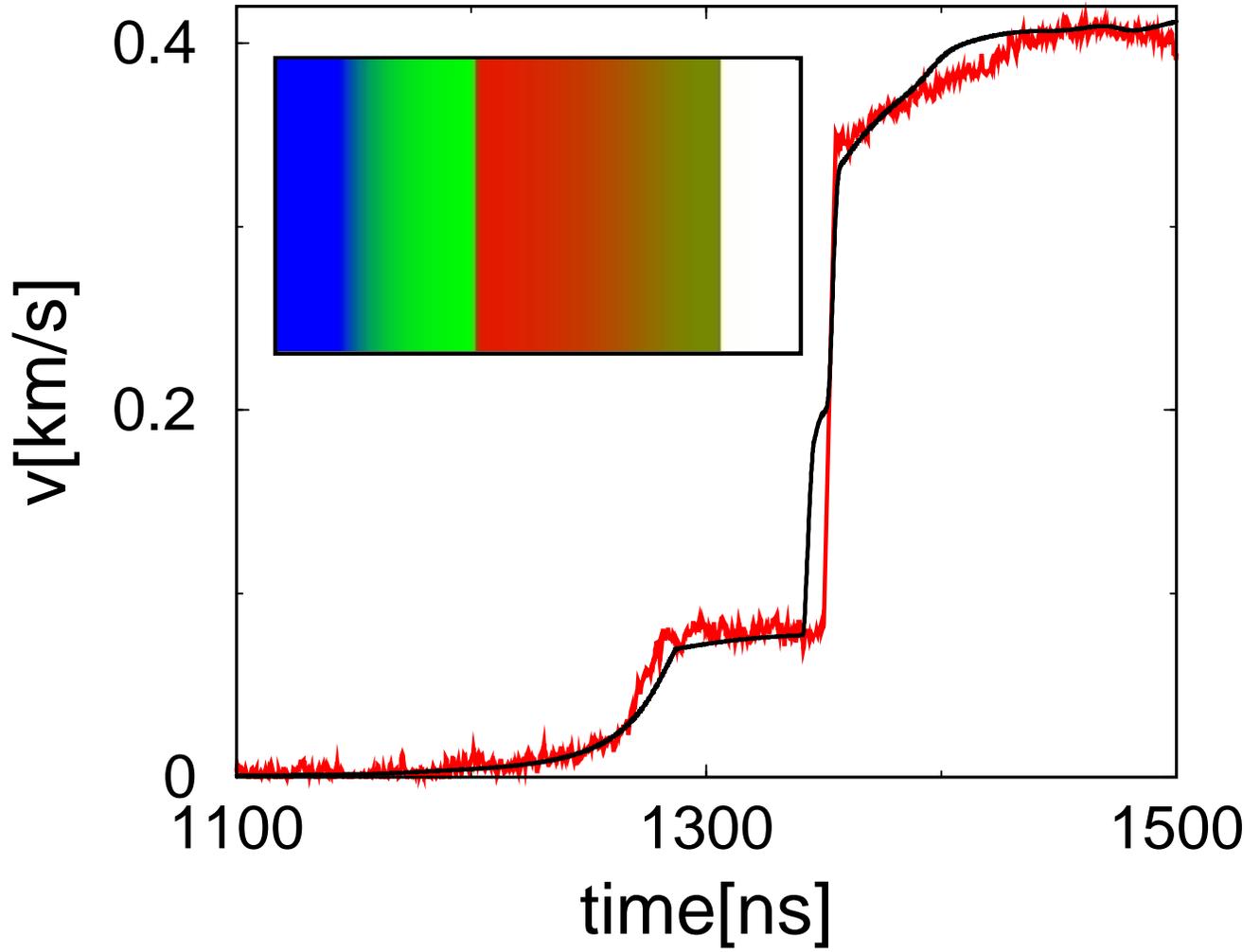}
\caption{VISAR trace (interface velocity) for the sapphire window experiment: red line; 
hydrodynamic simulations: black line. Inset: color-coded lateral cross section through the 
sample-window assembly showing the phase transformation fronts originating at the loading interface 
(left) and sapphire window (right) - Bi(I) (red), Bi(II)(green), Bi(III)(blue), sapphire (white) at 
$t\simeq 1310 ns$.}
\label{sapp}
\end{figure}

\begin{figure}
\includegraphics{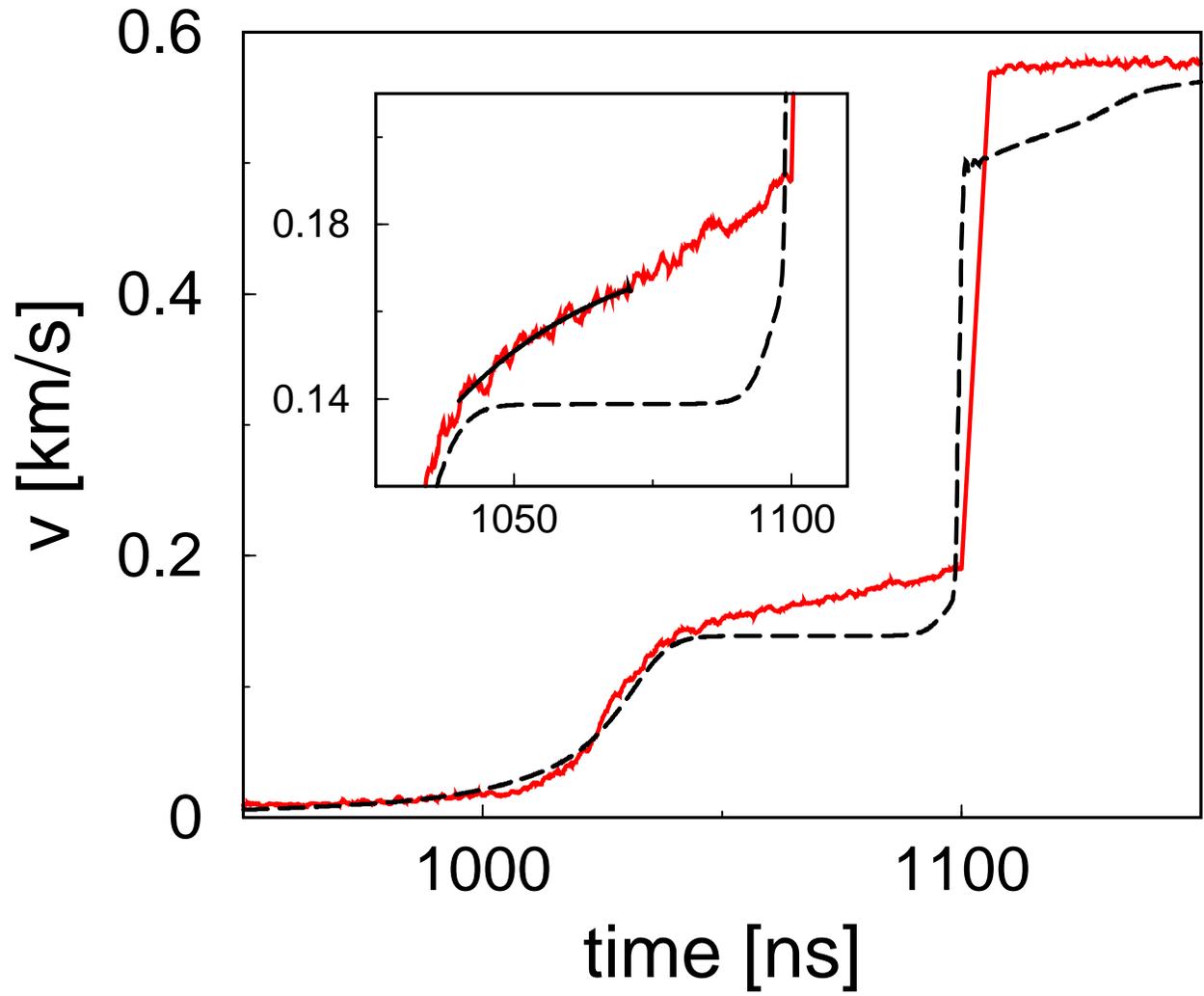}
\caption{VISAR trace (interface velocity) for a {\it LiF} window experiment: red line; 
hydrodynamic simulations: dashed black line. Inset: blow-up of the reduced acceleration 
regime and kinetics fit (see text): solid black line.}
\label{LiF}
\end{figure}

The behavior of bismuth during these experiments can be partly understood 
by comparing the sample/window interface velocity traces - $v(t)$, 
with the results of standard, equilibrium, one-dimensional hydrodynamic simulations. We 
performed such calculations using a multi-phase bismuth equation of state derived from 
the free energy model of Ref. \cite{hayes-Bi-EOS}, which describes very well the phases 
of main interest here - Bi(I) and Bi(II), and the position of their phase boundary; the 
higher pressure phase - Bi(III) is represented with lower accuracy, but that should not alter 
our conclusions. To insure accurate modeling the panels and windows were also included and 
described by Mie-Gruneisen equations of state \cite{Marsh}. The maximum densities achieved 
for bismuth were $\simeq 12.8 g/cm^3$ and the temperatures were below $\simeq 550K$.
The simulation results indicate that upon compression a structural phase 
transformation from the initial Bi(I) rhombohedral structure to the Bi(II) 
centered monoclinic structure (with a $5\%$ volume collapse) is initiated at $\simeq 19-24 
kbar$ and $\simeq 320-410K$, depending on the initial conditions. The transition is signaled 
in both experiments and simulations by a change in the slope of $v(t)$ - see Figs. \ref{sapp} 
and \ref{LiF}, which is followed in the simulations by a velocity ``plateau''; similar 
effects have been observed in shock experiments \cite{dg77}. Due to the 
complex wave interactions associated with the presence of material interfaces the pressure 
distribution inside the sample, and therefore the position dependent thermodynamic paths 
followed, are directly dependent on the compressive properties of the window. 

In the case of the sapphire window both the inception of the plateau and the measured 
velocity value, $\simeq 0.07km/s$, agree well with the hydrodynamic calculations, 
Fig. \ref{sapp}. A detailed analysis of the simulation results reveals that the 
transformation is initiated both at the loading-interface, due to the applied pressure, 
and at the back-interface, due to the pressure enhancement created by the ``hard'' 
(higher dynamic impedance) sapphire window, resulting in two transformation fronts 
traveling in opposite directions - see inset to Fig.\ref{sapp}. The start of the 
non-accelerating regime (velocity plateau) corresponds to the beginning of the phase 
transformation at the back-interface, while its end and the sharply rising velocity mark 
the completion of the transition in the entire sample.

For the case of the $LiF$ window on the other hand the experimental traces show 
surprisingly large deviations from the simulations. A softer window such as {\it LiF} creates 
a slight depressurization at the sample/window interface. Equilibrium hydrodynamic 
simulations indicate that the transformation front initiated at the loading-interface 
is traveling through the sample largely undisturbed by the back-interface. The velocity 
plateau starts when the perturbation generated by the advancing transformation reaches the interface, 
and is a thermodynamic equilibrium effect. No such plateau is observed in the 
experiment, although a marked change in acceleration, $\partial v/\partial t$, is present, 
see Fig. \ref{LiF}. This transient regime ends as before upon completion of the phase 
transition in the entire sample, as shown by the hydrodynamic calculations. 

In order to understand these results we consider the effect of solid-solid phase transition 
kinetics on dynamically driven phase transformations. In the present experiments 
bismuth undergoes a reconstructive structural first-order phase transformation, 
the kinetics of which can be described by a simple picture of nucleation and growth 
originally proposed by Kolmogorov \cite{k37}. This model is currently known as the 
Kolmogorov-Johnson-Mehl-Avrami (KJMA) model \cite{jm39,a41,c73}, and it has been 
employed to describe a variety of systems \cite{edkg94,kpng98,kg01,blct04}; 
we recall the main ideas below. For other, more detailed models of nucleation 
and growth see also \cite{loc92}.

If the system, initially in thermodynamic equilibrium in phase 1, is suddenly forced, e.g. 
by increasing the pressure, into the phase 2 region of its phase diagram, infinitesimally 
small domains of the stable phase will occur uniformly throughout the sample with a 
nucleation rate per unit volume $\gamma(t)$. Once formed the domains grow isotropically 
with constant interface velocity $u$, i.e. the rate of volume growth of a domain is assumed 
proportional with its surface area. At a later time $t$ the radius of a nucleus generated 
at $t\prime$  will be $r(t-t\prime)=u\times(t-t\prime)$, and its volume growth rate 
$w(t-t\prime)=4\pi u^3(t-t\prime)^2$. Therefore the unimpeded growth rate of the volume 
fraction of phase 2, $\phi_2$, will be $W(t)=\int_{0}^{t}w(t-t\prime)\gamma(t\prime)dt\prime$.
However, the growth of the 2-nd phase can only occur in the volume still occupied by the 
1-st phase, $1-\phi_2$, and as a result the actual growth rate is assumed proportional with $W$ 
and the volume still available. If at $t\rightarrow\infty$ the two phases coexist in thermodynamic 
equilibrium with volume fractions $\phi^0_1$ and $\phi^0_2$, 
the available volume is only $\phi^0_2-\phi_2$, and the rate of change of $\phi_2$ is: 
\begin{eqnarray}
\frac{\partial\phi_2}{\partial t}=(\phi^0_2-\phi_2)W(t)
\label{eq:kjma1}
\end{eqnarray}
This equation can be easily integrated if the system is not 
externally driven, e.g. by varying the applied pressure, i.e. $\phi^0_2$ is constant in time:
\begin{eqnarray}
\phi_2(t)=\phi^0_2\left\{1-\exp\left[-\int_{0}^{t}W(t\prime)dt\prime\right]\right\}
\end{eqnarray}

Two simple cases of the above equation have been often studied. One corresponds to 
time independent nucleation rate, also known as homogeneous nucleation. The other describes 
a situation where the nucleation process occurs primarily on defects, e.g. grain boundaries, 
or impurities already present in the sample, i.e. heterogeneous nucleation. In particular if 
the preexisting nucleation sites are assumed randomly distributed in the system with a number 
density $\gamma_0$ this formally corresponds to Eq. \ref{eq:kjma1} with 
$\gamma(t)=\gamma_0 \delta(t)$. In both cases Eq. \ref{eq:kjma1} reduces to:
\begin{eqnarray}
\phi_2(t)=\phi^0_2\left\{1-\exp{\left[-\left(\frac{t}{\tau}\right)^n\right]}\right\}
\label{eq:hohen}
\end{eqnarray}
where the kinetic time constant is $\tau\propto(\gamma u^3)^{-\frac{1}{4}}$ 
for homogeneous nucleation, and $\tau_0\propto(\gamma_0 u^3)^{-\frac{1}{3}}$ for 
heterogeneous nucleation. For the homogeneous case $n=4$ and $n=3$ for the heterogeneous one. 
However, $n$ can be interpreted more generally as a measure of the effective dimensionality 
of domain growth, which for heterogeneous nucleation in particular 
can be smaller than $3$, as first discussed by Cahn \cite{c56}. 

The modeling of the dynamic compression experiments described here 
requires the coupling of the transformation kinetics Eq. \ref{eq:kjma1} with appropriate 
macroscopic conservation equations for mass, momentum and energy, i.e. hydrodynamic 
equations. We now introduce a model and analysis that capture the effect of phase 
transformation kinetics on the propagation of perturbations through the 
system, and allow a quantitative interpretation of the experimental results.

Consider a semi-infinite sample in thermodynamic equilibrium at temperature $T$, coexistence 
pressure $P_c$ and density $\rho_1$ corresponding to the lower density phase. We are 
interested in the behavior of the system under a small perturbation, e.g. slight uniaxial 
compression with frequency $\nu$. If we neglect heat exchange processes, i.e. assume that 
the flow is isentropic, only mass and momentum conservation equations - Euler 
equations \cite{llfm} - need to be considered. Together with Eq. \ref{eq:kjma1} they 
constitute our coupled kinetics-hydrodynamics model. For small enough density and 
velocity deviations from equilibrium linearized Euler equations are sufficient, and read:
\begin{eqnarray}
\frac{\partial\rho}{\partial t} = -\rho_1\frac{\partial v}{\partial z}
\label{eq:linh1}
\end{eqnarray}
\begin{eqnarray}
\frac{\partial v}{\partial t} = -\frac{1}{\rho_1}\frac{\partial p}{\partial z}
\label{eq:linh2}
\end{eqnarray}
Eqs. \ref{eq:kjma1},\ref{eq:linh1} and \ref{eq:linh2} describe 
the propagation of small, long wavelength perturbations in the phase coexistence region of the phase diagram. 
To make further progress we use instead of Eq. \ref{eq:kjma1} the integrated form 
Eq. \ref{eq:hohen}, which should not introduce large errors since we expect that $\phi^0_2$ 
is a slow, hydrodynamic variable, which changes on time scales of order 
$\nu^{-1}$ ($\nu^{-1}\simeq500 ns$ in the experiment), much larger than the characteristic time 
scale of Eq. \ref{eq:kjma1}, i.e. $\nu^{-1}\gg \tau$. As 
usual this set of equations needs to be closed by expressing the pressure $p$ as a function 
of density $\rho$ and volume fraction $\phi_2$, as well as $\phi^0_2$ as function of $\rho$, 
all at constant entropy. We assume here that thermal as well as mechanical 
equilibrium prevail on time scales much shorter than $\tau$ in microscopically large but macroscopically 
small sample regions, for arbitrary local volume fractions $\phi_2$ of the coexisting phases. 
We also introduce further simplifying assumptions, for example that the differences between the 
densities and compressibilities of the two phases are 
small - e.g. they are $\simeq 5\%$ for bismuth I and II, and also that differences between 
the isentrope and an average isotherm are small, which holds well for bismuth at the typical 
experimental pressures and temperatures. We obtain for the velocity equation:
\begin{eqnarray}
\frac{\partial v}{\partial t} = -\frac{\chi_{1}}{\rho_1^2}\left(\frac{\partial\rho}
{\partial z}\right)
exp{\left[-\left(\frac{t-t_0}{\tau}\right)^n\right]}
\label{eq:linh3}
\end{eqnarray}
$t\geq t_0$, where $\chi_{1}$ is the adiabatic compressibility of phase 1 and $t_0$ is the 
arrival time of the compressive perturbation at position $z$. In conjunction with 
Eq. \ref{eq:linh1} the above relation yields a modified sound equation for the density. 
Guided by the experimental set-up, where the compression starts below the transition line, we argue 
that the density variations propagate approximately 
as sound waves with the frequency $\nu$ of the applied perturbation. 
Since we assume $\nu^{-1}\gg \tau$, we can therefore write for the velocity equation:
\begin{eqnarray}
\frac{\partial v}{\partial t} \simeq A(z) exp{\left[-\left(\frac{t-t_0}{\tau}\right)^n\right]}
\label{eq:veli}
\end{eqnarray}
$t\geq t_0$, where $A(z)$ (defined by comparison with Eq. \ref{eq:linh3}) is now time independent. 
For reasonably short time intervals $\delta t=t-t_0$ this equation should approximately govern the 
evolution of the velocity not too far ahead of the transformation front. 

We expect the above analysis, corresponding to a propagating transformation front in the 
vicinity of the phase line, to be suitable for the ``soft'' {\it LiF} window experiments. For 
this case we would therefore like to fit the experimentally measured back-interface velocity 
with the functional form Eq. \ref{eq:veli}, to obtain information on the effective kinetic 
time constant $\tau$ and the Avrami exponent $n$. To this end we set $t_0$ by comparing with 
instantaneous kinetics hydrodynamic calculations and restrict the fit to approximately one-half 
of the duration of 
the reduced acceleration regime, to avoid the effects of pressure reverberation between the 
transformation front and the {\it LiF} window. Coincidentally, the fit termination point 
can also be identified as an inflection point. A typical result is shown in Fig. \ref{LiF}, 
where we find $\tau\simeq 24ns$. This reasonably validates {\it {a posteriori}} 
the assumption of a scale separation between $\tau$ and $\nu^{-1}$. We determine an 
Avrami exponent $n\simeq 1.3$, which suggests strongly heterogeneous nucleation dominated by 
a high density of sites located on grain interfaces \cite{c56}. This is consistent with the 
poly-crystalline character of the bismuth samples used in the present experiments. For the 
``hard'' sapphire window on the other hand, an additional transformation front is generated 
at the sample/window interface due to pressure enhancement at that boundary. This occurs 
before the arrival of sound waves from the forward moving transformation front and thereby 
obscures its effect. 

As shown before, in the case of heterogeneous nucleation the characteristic time constant 
$\tau$ depends on the density of 
defects $\gamma_0$ and the phase interface velocity $u$; $\gamma_0$ is directly related to the 
average size of the grains for the case of grain boundary nucleation, while the interface 
motion is driven by both thermodynamical forces - the difference between the chemical 
potential of the two phases $\Delta \mu$, and mechanical ones - the applied loads and the 
elastic stresses that occur at the boundary between the competing phases due to their 
different densities and lattice structures \cite{fr98,vil02}. In the vicinity of the phase line the 
thermodynamical contribution to $u$ has a fairly simple form \cite{dt56,lai02}, 
$u\propto \lambda \Delta \mu \exp(-Q/k_BT)$, where $\lambda$ is the interface 
thickness and $Q$ the activation energy for atomic cross-interface motion; here $\Delta\mu$ 
should be interpreted as a time-averaged chemical potential difference. If we neglect 
the pressure dependence of $\lambda$ and $\Delta\mu$ (the phase line is to a good 
approximation flat) and assume that the exponential term contains the dominant 
temperature contribution, for similar samples crossing the phase line at different 
thermodynamic points the time constants $\tau$ should reflect an Arrhenius-type 
temperature dependence of the interface velocity. 
For {\it LiF} window experiments at transition temperatures $T_1\simeq 320K$, 
$T_2\simeq 360K$ and $T_3\simeq 410K$ we find $\tau$'s consistent 
with such a behavior, and an activation energy $Q\simeq 0.2eV$. 
Although the interplay between thermodynamical and mechanical forces is rather 
complex \cite{lhsa04}, this suggests that the thermodynamic force is dominant, at least 
in the initial stages of transformation kinetics. 

We believe that our experimental results and analysis 
provide new insight on phase transformation kinetics occurring under dynamic conditions, 
and open the possibility of experimentally designing and characterizing both thermodynamic 
and kinetic paths.

This work was performed under the auspices of the U. S. Department of Energy by 
University of California Lawrence Livermore National Laboratory under Contract 
No. W-7405-Eng-48.

\newpage

\newpage

\newpage

\end{document}